\documentclass[%
 reprint,
 amsmath,amssymb,
 aps,
]{revtex4-2}

\usepackage{graphicx}
\usepackage{dcolumn}
\usepackage{bm}
\usepackage{float} 
\usepackage{amsmath}
\usepackage{upgreek}
\usepackage{soul,xcolor}
\newcommand{\ket}[1]{|#1\rangle}

\usepackage[normalem]{ulem}

\begin{document}
\setstcolor{red}
\preprint{APS/123-QED}

\title{Real- and Fourier-space observation of the anomalous $\pi$-mode in Floquet engineered  plasmonic waveguide arrays}

\author{Anna Sidorenko}
 \email{a\textunderscore sidorenko@uni-bonn.de}
\author{Zlata Fedorova (Cherpakova)}%
 \email{cherpakova@physik.uni-bonn.de}
 \author{Johann Kroha}
 \email{kroha@th.physik.uni-bonn.de}
\author{Stefan Linden}
 \email{linden@physik.uni-bonn.de}
\affiliation{Physikalisches Institut, Rheinische Friedrich-Wilhelms-Universit\"at Bonn, Nußallee 12, 53115 Bonn, Germany
}%

\date{\today}

\begin{abstract}
We present a joint experimental and theoretical study of the driven Su-Schrieffer-Heeger model implemented by arrays of evanescently coupled plasmonic waveguides. Floquet theory predicts that this system hosts for suitable driving frequencies a topologically protected edge state that has no counterpart in static systems, the so-called anomalous Floquet topological $\pi$-mode. By using real- and Fourier-space leakage radiation microscopy in combination with edge- and bulk excitation, we unequivocally identify the anomalous Floquet topological  $\pi$-mode and study its frequency dependence. 
\end{abstract}

\maketitle

Floquet engineering, i.e., the time-periodic perturbation of an otherwise static lattice system by an external drive, is a powerful method to create and study topological phases of matter \cite{oka2009photovoltaic, lindner2011floquet, kitagawa2011transport,gomez2013floquet,cayssol2013floquet}.
The underlying idea  is that the modulation dynamically couples and hybridizes eigenstates of the static system whose eigenfrequencies differ by multiples of the modulation frequency, and by that modifies the spectrum of the system.
By choosing proper driving conditions, the resulting Floquet band structure can feature a qualitatively different character than the original band structure \cite{nathan2015topological}. 
In particular, the modulated system can possess nontrivial topological properties even though the underlying static system is trivial \cite{kitagawa2010topological,PhysRevX.3.031005,rechtsman2013photonic}.
What is even more remarkable, topological insulators created by time-periodic driving, so-called Floquet topological insulators, can show topological properties that have no complement in static systems.
For instance, periodically driven, two-dimensional lattices can feature chiral edge states despite vanishing Chern numbers of the bulk Floquet bands~\cite{kitagawa2010topological,PhysRevX.3.031005,maczewsky2017observation}. 

The Su-Schrieffer-Heeger (SSH) model~\cite{SU1997}, i.e., a one-dimensional tight-binding lattice with staggered hopping amplitudes $J_1$ and $J_2$, has attracted considerable interest as a simple yet topologically non-trivial system~\cite{asboth2016short}. 
For static hop\textcolor{violet}{p}ing amplitudes, its spectrum consists of two bands separated by a gap of width $2\vert J_1-J_2\vert$.
Depending on the choice of the unit cell (either $J_1 > J_2$ or $J_1<J_2$), the model features two dimerizations with different Zak-phases~\cite{delplace2011zak}. In virtue of chiral symmetry, these dimerizations correspond to two topologically distinct phases. As a consequence of the bulk-boundary correspondence principle, each interface between two domains of different topology supports a topologically protected edge state with zero energy.

Periodic modulation of the hopping amplitudes splits the original band structure of the undriven SSH model into infinitely many equivalent copies (Floquet replicas) spaced by the driving frequency $\omega$ (here and in the following we set $\hbar=1$)~\cite{gomez2013floquet,dal2015floquet}.
In the high frequency regime, band gaps between the Floquet replicas trivially arise since $\omega$ is larger than the widths of the bands. Driving induced gaps can, however, also evolve at lower frequencies for which the Floquet replicas spectrally overlap. In this case, the modulation  hybridizes states at the boundary of the Floquet Brillouin zone and hence lifts their degeneracy~\cite{gomez2013floquet,dal2015floquet}.
For suitable driving frequencies, if the topological invariants of the hybridizing bands have different values, the resulting gap can host a topologically protected edge state, the so-called anomalous Floquet topological $\pi$-mode. 
In contrast to the zero-energy mode of the undriven SSH model, it has a quasienergy $E=\pm \omega/2$,  i.e., it is located at the boundary of the first Floquet Brillouin zone.
Note, that such modes can only arise in driven systems because the band gaps supporting them are due to periodicity in quasienergy of the Floquet spectrum. 
Following theoretical predictions~\cite{asboth2014chiral,fruchart2016complex,dal2015floquet,zhang2017edge}, the anomalous Floquet
topological $\pi$-mode has recently been observed in real space by microwave near-field experiments performed on a metallic array of coupled corrugated waveguides~\cite{cheng2019observation} and at optical frequencies in a non-Hermitian waveguide lattice~\cite{wu2021floquet}.

\begin{figure}[t]
\centering
\includegraphics{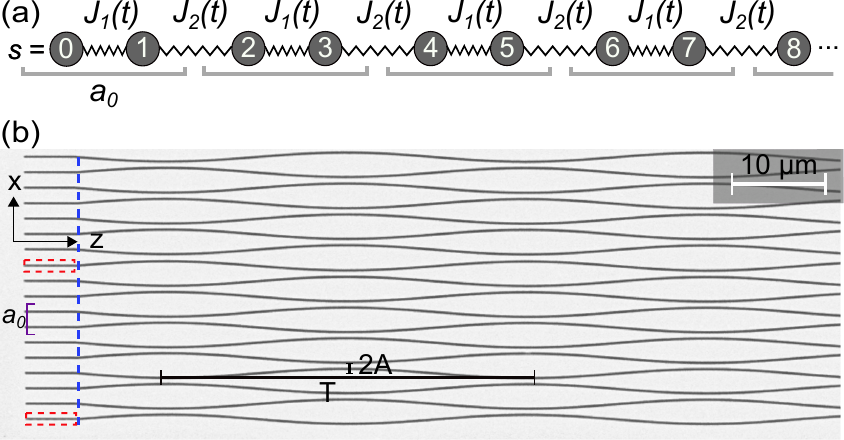}
\caption{\label{Fig1_label} (a) Sketch of the driven SSH model with time-periodic coupling constants $J_1(t)$ and $J_2(t)$ modulated with frequency $\omega$. (b) Scanning electron micrograph of a plasmonic waveguide array used to implement the driven SSH model. The blue dashed line marks the boundary of the excitation region and the red dotted boxes indicate the position of the grating couplers deposited onto the input waveguides.}
\end{figure}

 In this work, we present a joint experimental and theoretical study of the driven SSH model with time-periodic coupling constants. We unambiguously identify the anomalous Floquet
topological $\pi$-mode by its localization at the system boundary in real space as well as spectroscopically by its characteristic energy of half the driving frequency, $\pm E=\omega/2$.
Our experimental implementation of the model is based on 
evanescently coupled dielectric-loaded surface-plasmon polariton waveguide arrays (DLSPPWs) operating at near-infrared frequencies (see Fig.\,\ref{Fig1_label}). 
Based on the mathematical equivalence of the paraxial Helmholtz equation and the time-dependent Schr\"odinger equation~\cite{christodoulides2003discretizing,longhi2009quantum}, the spatial surface plasmon polariton (SPP) evolution in such a waveguide array can be mapped onto the temporal dynamics of an electronic wave packet in the corresponding one-dimensional SSH model.
Effects of different driving frequencies are studied using leakage radiation microscopy (LRM)~\cite{drezet2008leakage} as detection technique, which gives us the opportunity to not only observe the intensity distribution of the modes in real space but also image their momentum-resolved spectra. 
The experimental findings are supported by numerical calculations based on the Floquet formalism.

\begin{figure}[tttt]
	\centering
	\includegraphics{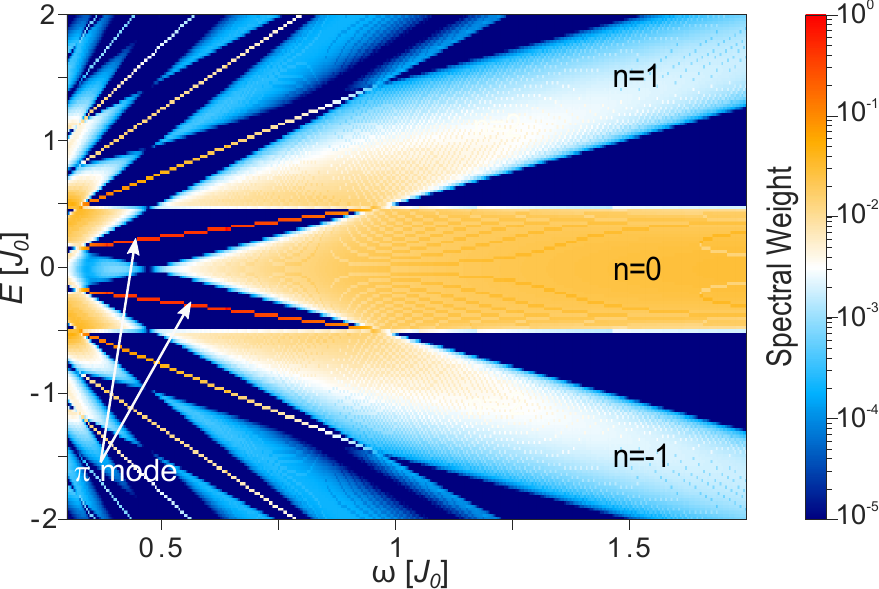}
	\caption{\label{Fig2} Calculated frequency-dependent quasienergy spectrum with assigned weights in case of single-site excitation at the edge. Calculations were performed for $N=50$ unit cells. 
	}
\end{figure}

In the following, we consider the periodically driven SSH model with time-dependent coupling constants (see Fig.\,\ref{Fig1_label} (a)). The corresponding Hamiltonian can be written as~\cite{asboth2016short}:  
\begin{equation}
\label{eq:hamiltonian}
\hat {H}(t)=\sum_s J_1(t) \hat {a}_{2s}^\dagger \hat {a}^{\phantom{\dagger}}_{2s+1}+J_2(t) \hat {a}_{2s+1}^\dagger \hat {a}^{\phantom{\dagger}}_{2s+2}+h.c.,
\end{equation}
where $\hat {a}_{s}^\dagger$ ($\hat{a}^{\phantom{\dagger}}_{s}$) is the creation (annihilation) operator acting at the lattice site $s$. The inter-/intra-cell hoping amplitudes $J_{1/2}(t)$ are real-valued, periodic functions of time with frequency $\omega = 2\pi/T$, where $T$ is the period of driving. 
In the fabricated samples, we sinusoidally vary the  separation between neighboring waveguides (see Fig.\,\ref{Fig1_label} (b)).
Since the mode overlap has an exponential dependence on the spacing between neighboring sites, we write the hopping amplitudes as:
\begin{subequations}
\begin{align}
    \label{eq:design}
    J_1(t)&=J_0 e^{-\lambda \left[1-\sin({\omega t})\right]}, \\
    J_2(t)&=J_1 (t-T/2),
\end{align}
\end{subequations}
where $\lambda$ characterizes the exponentially decaying mode overlap.
This coupling scheme, where the two coupling amplitudes have the same period-averaged value, has been chosen in order to close the band gap at zero quasienergy (see  Fig.~\ref{Fig2}). In this way, we intentionally eliminate the zero-energy topological edge state also found in the static SSH model.

The Hamiltonian of the driven SSH model is periodic in time, $\hat{H}(t+T)=\hat{H}(t)$ so that the Floquet theory lends itself to the theoretical analysis ~\cite{shirley1965solution,gomez2013floquet,holthaus2015floquet,eckardt2017colloquium,gomez2013floquet}. 

According to Floquet's theorem, solutions of the time-dependent Schr\"odinger equation, $i\frac{\partial }{\partial t}\ket{\psi(t)}=H(t)\ket{\psi (t)}$, can be expanded in a set of Floquet states~\cite{gomez2013floquet}:
\begin{equation}
    \label{eq:Floquet}
    \ket{\psi_\alpha(t)} = \exp(-i\epsilon_{\alpha}t)
    \ket{u_{\alpha}(t)}.
\end{equation}
where $\epsilon_{\alpha}$ is the quasienergy of the state, $\ket{u_{\alpha}(t+T)}=\ket{u_{\alpha}(t)}$ is the associated time-periodic Floquet mode, and $\alpha$ is the mode index. 
The quasienergies are determined only up to integer multiples of $\omega$, hence,  the $\epsilon_{\alpha}$ restricted to the first Floquet Brillouin zone (FFBZ) $[-\omega/2,\omega/2 )$ define all unique Floquet states.
As immediately follows from the time-dependent Schr\"odinger equation, the quasienergies are solutions of the following eigenvalue problem:
\begin{equation}~\label{eq:EgenvalueProblem}
\left(\hat{H}(t)-i\frac{\partial}{\partial t}\right)\ket{ u_\alpha(t)}=\epsilon_\alpha\ket{u_\alpha(t)}.    
\end{equation}
Due to temporal periodicity of the Hamiltonian and the Floquet modes, they  can be expanded into Fourier series with $\hat{H}(t)=\sum_n e^{-\imath n \omega t}H_n$ and $\ket{u_{\alpha}(t)}=\sum_n e^{-\imath n \omega t}\ket{u_{\alpha}^n}$, respectively. Substitution of these series into Eq.~\eqref{eq:EgenvalueProblem} yields the time-independent Floquet equation
\begin{equation}~\label{eq:TimeIndependentFloquet}
(H_0-\hbar\omega)\vert u_\alpha^n\rangle+\sum_{m\neq 0}H_m\vert u_\alpha^{n-m}\rangle = \epsilon_\alpha \vert u_\alpha^n\rangle   
\end{equation}
for all integer Floquet indices $n,m$. As a result, our 1-dimensional time-periodic system is represented as a (1+1)-dimensional time-independent system whose spectrum is composed of infinitely many equivalent copies (Floquet replicas) described by $H_0$ and coupled with each other by means of $H_m$, $m\neq0$.
Eq.~\eqref{eq:TimeIndependentFloquet} can be solved numerically by introducing a cutoff index $m'$ such that $\ket{u_\alpha^n}=0$ for $\vert n\vert > m'$~\cite{fedorova2021dissipation}.

The complete solution of the Schr\"odinger equation can be written as 
\begin{equation}
    \label{eq:Floquetfull}
    \ket{\Psi(t)} = \sum_\alpha C_\alpha \sum_n \exp(-i\epsilon_{\alpha}^n t)
    \ket{u_{\alpha}^n},
\end{equation}
with $\epsilon_{\alpha}^n=\epsilon_{\alpha}+n \omega$.
The constants $C_{\alpha}$ depend on the initial condition and can be calculated as $C_{\alpha}=\langle u_{\alpha}(0)|\Psi(0)\rangle$. The temporal Fourier transform of equation~(\ref{eq:Floquetfull}) yields $\ket{\psi(E)}=\sum_{\alpha,n}C_\alpha \ket{u_{\alpha}^n}\delta(E-\epsilon_{\alpha}^n)$. 
Therefore, the normalized spectral weight at quasienergy $E=\epsilon_{\alpha}^n$ is given by:
\begin{equation}
    \label{eq:weight}
    \vert\psi (\epsilon_{\alpha}^n)\vert^2 = \left|C_{\alpha}\right|^2 \langle u_{\alpha}^n|u_{\alpha}^n\rangle.
\end{equation}
Figure~\ref{Fig2} depicts the calculated spectral weights of the wavefunction injected at $t=0$ at the outermost site ($s=0$) in dependence on the quasienergy and the driving frequency from the experimentally achievable range $\omega\in [0.3 J_0,1.75 J_0]$. Here, we assume a lattice size of $N=50$ unit cells, $J_0=1$, $\lambda = 3.02$, and all energies are given in units of $J_0$, similar to~\cite{fedorova2020observation}.

The quasienergies $\epsilon_{\alpha}$ were obtained by solving the truncated version of Eq.~\eqref{eq:TimeIndependentFloquet}. The
color coding in Fig.~\ref{Fig2} indicates the spectral weight from Eq.~\eqref{eq:weight}, calculated for a state $\ket{\Psi(t)}$ [Eq.~\eqref{eq:Floquetfull}], with initial conditions chosen such that the bulk and the  spatially localized edge states can be populated simultaneously.

We start our discussion with the high-frequency regime ($\omega>1.0 J_0$).

As mentioned above, the spectrum has no gap at zero quasienergy for the chosen driving scheme (see Fig.~\ref{Fig2}). Instead, we observe gaps between the different Floquet replicas of the bands formed by the bulk modes, since the driving frequency is larger than the width of the bands.
In the following we concentrate on the so-called $\pi$ gap between the $n=0$ and $n=1$ replicas.
In accordance with the previous calculations~\cite{dal2015floquet,zhang2017edge}, we observe in the high frequency regime no midgap state and, hence,  can conclude that the system is topologically trivial in this case. 

When lowering the driving frequency, 
the $\pi$-gap opens and closes repeatedly.
This occurs whenever a new pair of Floquet replicas gets into or out of resonance at the boundary of the Floquet Brillouin zone.
As the Zak phase acquires a $\pi$-phase shift on this occasion, the $\pi$–gap repeatedly switches its topological character~\cite{gomez2013floquet,dal2015floquet}. 
In particular, between the first and the second gap closing points ($0.32 J_0<\omega< 1.0 J_0$), the $\pi$-gap becomes topologically nontrivial and hosts a mid-gap state (white  arrows in Fig.~\ref{Fig2}), i.e., the anomalous Floquet topological $\pi$-mode.
If we excite the system in the middle of the array, i.e., in the bulk, the anomalous Floquet topological $\pi$-mode is absent in the spectrum  (not shown). 
This is a consequence of the exponential localization of the anomalous Floquet topological  $\pi$-mode at the edge (see below).
For specific frequencies, e.g., $\omega=0.47 J_0$, the widths of the bands collapse almost completely and we expect dynamic localization of wavepackets also in the bulk~\cite{holthaus1992collapse,zhang2017edge}. We note that this effect is not of topological nature but a consequence of the vanishing group velocity for these specific frequencies~\cite{zhang2017edge,fedorova2021dissipation}. 

\begin{figure*}[ttt]
	\centering
	\includegraphics[width=\textwidth]{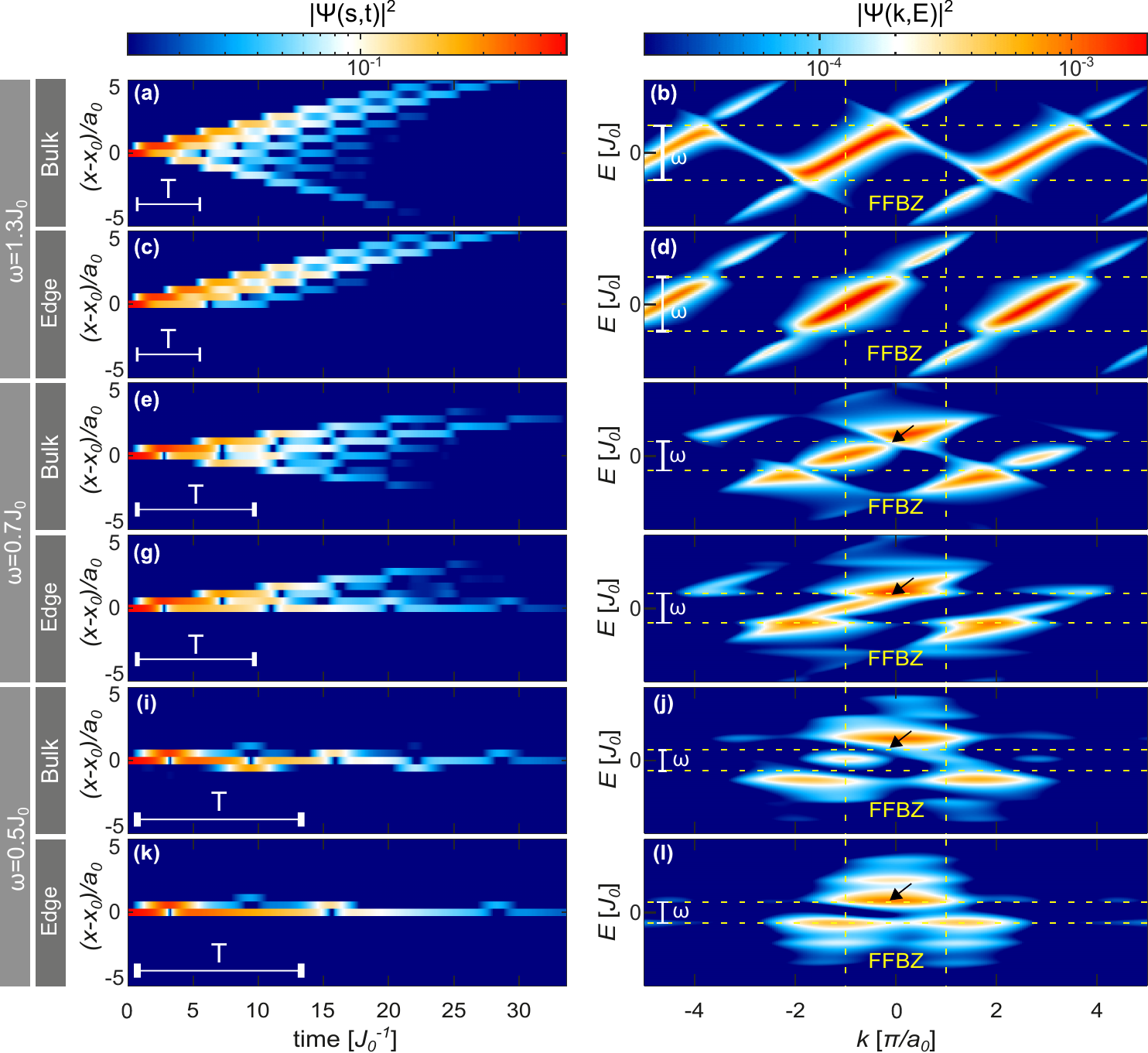}
	\caption{\label{fig:theory2} Calculated temporal evolution of the probability density $\vert\psi(s,t)\vert^2$ (left) and corresponding momentum resolved spectra $\vert\psi(k,E)\vert^2$ (right) for different driving frequencies and single-site excitation at the position $x_0$ either in the bulk or the edge lattice site. All  calculations were performed for the periodically driven SSH model with $N=50$ unit cells. (\textbf{a})- (\textbf{d}) High frequency regime ($\omega$ = 1.3$J_0$). (\textbf{e})-(\textbf{h}) - Intermediate frequency regime ($\omega$ =0.7$J_0$). (\textbf{i})-(\textbf{l}) Low frequency regime, dynamic localization ($\omega$ = 0.5$J_0$). The white scale bars in the real-space data indicate the corresponding driving period $T$ while the scale bars in the spectra show $2\omega$. In the spectra excited in the bulk, the arrows point to the locations of the band gap, while for the edge excitation they highlight the anomalous $\pi$-mode.  
	}
\end{figure*}

The left-hand side of Fig.~\ref{fig:theory2} depicts the calculated temporal evolution of the probability density $\vert\psi(s,t)\vert^2$, where $\psi(s,t)$ is the projection of $\vert \psi(t)\rangle$ onto the lattice sites $s$, 
for three characteristic driving frequencies and single site excitation either in the bulk ($x_0=N/2$) or at the edge ($x_0=0$) of the lattice.
The absolute value squared of the 2D Fourier transform $\vert\psi(k,E)\vert^2$ yields the corresponding momentum-resolved spectra plotted on the right-hand side. 
Based on auxiliary finite element calculations,  we assume a global damping rate of $\gamma=0.098J_0$ for all modes to account for the ohmic losses inherent for propagation of SPPs. 
Note, that such homogeneous losses only cause the exponential decay of the probability density in real space and spectral line broadening in Fourier space but otherwise do not alter the systems dynamics.
For $\omega=1.3 J_0$ and excitation in the bulk (see Fig.\,\ref{fig:theory2}(a)), the temporal evolution of the probability density is governed by  ballistic spreading of the wave packet. The  asymmetry with respect to the $x$-axis is a result of the initial condition that effectively breaks space and time inversion symmetries~\cite{fedorova2021dissipation}. The corresponding momentum-resolved spectrum is composed of a set of Floquet bands with almost linear dispersion and band gaps at the Floquet Brillouin
zone boundaries (see Fig.\,\ref{fig:theory2}(b)). 
Excitation at the edge of the lattice also leads to ballistic spreading of the wave packet over time without an indication of localization at the boundary (see Fig.\,\ref{fig:theory2}(c)).
Since we excite in this case exclusively states moving in $+x$-direction, the momentum-resolved spectrum  contains only Floquet bands with positive slope (see Fig.\,\ref{fig:theory2}(d)).

If the driving frequency is lowered down to $\omega=0.7 J_0$ and the system is excited in the bulk, the temporal evolution of the probability density (see Fig.\,\ref{fig:theory2}(e)) and the momentum resolved spectrum (see Fig.\,\ref{fig:theory2}(f)) show similar characteristics as in the high frequency case discussed above. In contrast, excitation at the edge leads to a qualitatively new feature for this driving frequency.
The temporal evolution of the probability density shows that a large fraction of the excited population remains localized to the edge and is periodically exchanged between the two outermost sites (see Fig.\,\ref{fig:theory2}(g)). In the corresponding momentum resolved spectrum, new modes with quasienergies $E=\omega/2\pm n\omega$ appear as bright horizontal lines in the middle of the band gap (see Fig.\,\ref{fig:theory2}(h), magenta arrow).
This is exactly the expected signature of the anomalous Floquet topological $\pi$-mode~\cite{cheng2019observation}.  

For $\omega=0.5 J_0$ and bulk excitation, the momentum resolved spectrum is composed of nearly flat bands (see Fig.\,\ref{fig:theory2}(j). In the time domain (see Fig.\,\ref{fig:theory2}(i)), this is connected with a very slow spreading of the wave packet, i.e., dynamic localization. In the case of edge excitation, we again observe the anomalous Floquet topological $\pi$-mode identified by time-periodic oscillations of the population at the edge of the lattice (see Fig.\,\ref{fig:theory2}(k)) and its spectral position at the boarder of the Floquet Brillouin zone (see Fig.\,\ref{fig:theory2}(l)). Thus, at this particular frequency the spectrum consists of equidistant horizontal lines separated by $\omega$ no mater whether the excitation takes place at the edge or in the bulk.

In order to test the theoretical predictions, we perform experiments on arrays of dielectric-loaded surface plasmon-polariton waveguides (DLSPPWs).
Based on the quantum-optical analogy~\cite{christodoulides2003discretizing,longhi2009quantum}, the propagation distance along the waveguides ($z$-axis) takes the role of time. Accordingly, the real-space SPP intensity $I(x,z)$ corresponds to the probability density $\vert\psi(x,t)\vert^2$ and the Fourier-space SPP intensity $I(k_x,k_z)$ correlates with the  spectrum $\vert\psi(k,E)\vert^2$ decomposed in the momentum components in the different Brillouin zones~\cite{bleckmann2017spectral}.

\begin{figure*}[ttt]
\centering
\includegraphics[width=\textwidth]{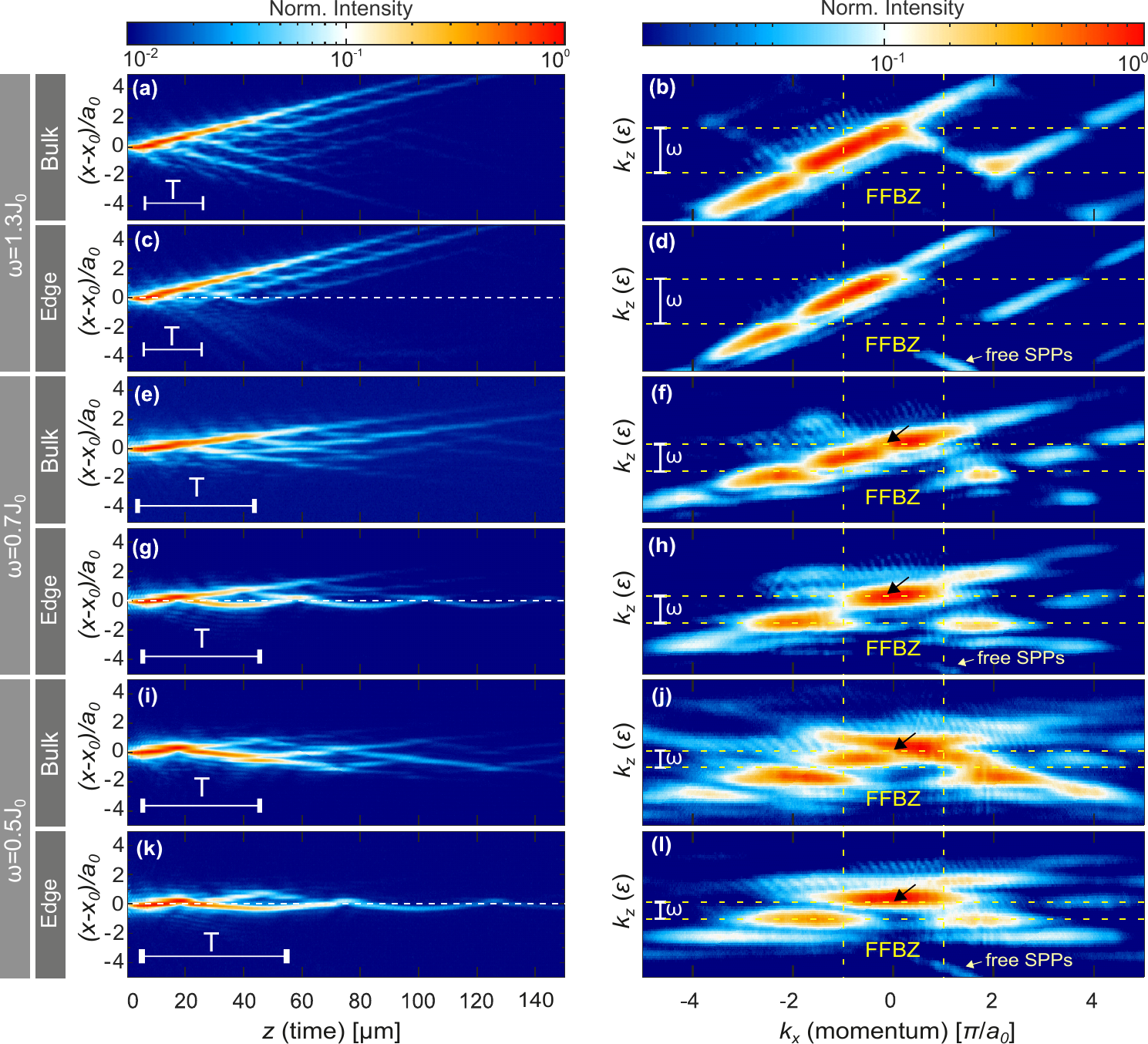}
\caption{\label{Fig4label} Measured real-space (left) and Fourier-space (right) SPP intensity distributions for different driving frequencies and single-site excitation at the position $x_0$ either in the bulk or the edge waveguide. (\textbf{a})-, (\textbf{d}) High frequency regime (T = 20 $\upmu$m corresponding frequency $\omega$ = 1.3$J_1$). (\textbf{e})-(\textbf{h}) Intermediate frequency regime (T = 40 $\upmu$m corresponding frequency $\omega$ =0.7$J_1$). (\textbf{i})-(\textbf{l}) Low frequency regime (T = 50 $\upmu$m corresponding frequency $\omega$ = 0.5$J_1$). The white scale bars in the real-space data indicate the corresponding driving period $T$ while the scale bars in the spectra show $2\omega$. In the Fourier-space images, the arrows indicate the predicted band gaps for excitation in the bulk, while for the edge excitation they highlight the location of the anomalous $\pi$-mode. }
\end{figure*}

The samples were fabricated by negative-tone grey-scale electron beam lithography~\cite{bleckmann2013manipulation,block2014bloch} and consist of arrays of poly(methyl methacrylate) (PMMA) ridges deposited on a microscope cover slip covered with $5\,\mathrm{nm}$ chromium (adhesion layer) and $60\,\mathrm{nm}$ gold. Fig.~\ref{Fig1_label} (b) shows a micrograph of a typical waveguide array used in our experiments. Width and height of the waveguides are $350\,\mathrm{nm}$ and $130\,\mathrm{nm}$, respectively. 
The periodic modulation of the coupling constants is implemented by sinusoidally  varying  the  center-to-center distance between  neighboring  waveguides.
The mean center-to-center distance between adjacent waveguides and the modulation amplitude are chosen to be $d=1.7 \,\upmu\mathrm{m}$ and $A=0.5\,\upmu\mathrm{m}$, respectively. 
We note that the variation of the effective refractive index due to the curvature of the waveguides can be be neglected because of the strong confinement of the SPPs in the waveguides, i.e., we can set the on-site potential $V_{0}(t) \approx 0$.
Combining an auxiliary experiment with two waveguides (not shown) and finite element calculations, we determine that the coupling constants vary between $J_0=0.24$ $\upmu \text m^{-1}$ and 
0.0006\ $\upmu \text m^{-1}$ for these parameters. 
Different driving frequencies are realised by varying the modulation period $T$. 
For the chosen average center-to-center distance and modulation amplitude, the modulation periods $T=20\,\upmu\mathrm{m}$, $T=40\,\upmu\mathrm{m}$ and $T=50\,\upmu\mathrm{m}$ correspond to the driving frequencies considered above in the calculations, i.e., $\omega=1.3 J_0$, $\omega=0.7 J_0$, and $\omega=0.5 J_0$, respectively. 
A short section of straight waveguides in front of the modulated part serves as an excitation region (part on the left hand-side of the blue vertical line in Fig.~\ref{Fig1_label} (b)) that contains the grating couplers deposited on the outermost waveguide and on a waveguide in the interior of the array (see red boxes in Fig.~\ref{Fig1_label} (b)).

Real- and Fourier-space images of the SPP intensity distributions are recorded by leakage radiation microscopy ~\cite{drezet2008leakage}. For this purpose, SPPs are excited by focusing a TM-polarized laser beam with $\lambda_0 = 980\,\mathrm{nm}$ wavelength onto the chosen grating coupler. An oil immersion objective (Nikon 60$\times$, numerical aperture NA = 1.4 Plan-Apo) is used to collect the leakage radiation as well as the transmitted laser beam. The latter is filtered out by a knife edge placed in the intermediate back focal plane (BFP) of the objective. The leakage radiation is imaged onto an sCMOS camera (Andor Marana). The real-space SPPs intensity distribution is recorded at the real image plane while the Fourier-space intensity distribution is acquired by imaging the BFP of the oil immersion objective.

The measured real-space and Fourier-space SPP intensity distributions presented in Fig.~\ref{Fig4label} qualitatively confirm the predictions of the numerical calculations. 
In the high frequency case ($\omega=1.3 J_0$),
we observe, regardless of the excitation position, ballistic spreading of the wave packet in the real-space intensity distributions (bulk: Fig.~\ref{Fig4label} (a); edge: Fig.~\ref{Fig4label} (c)) without an indication of localization. As discussed above, the initial conditions lead to an imbalance between modes propagating in $+x$- and $-x$-direction. The main feature of the related Fourier-space distributions, which correspond to the momentum resolved spectra, are sets of bands with nearly linear dispersion separated by band gaps (bulk: Fig.~\ref{Fig4label} (b); edge: Fig.~\ref{Fig4label} (d)). We note that in any realistic experiment, in addition to ohmic losses, the propagation of SPPs is inevitably associated with extra losses due to leakage radiation and imperfections of the metal film. 
As a result, the bands show additional line broadening  in comparison to the numerical calculations shown on the right-hand side of Fig.~\ref{fig:theory2}.
We note in passing that the curved spectral feature in the low $k_z$-region can be attributed to the excitation of free SPPs on the bare gold film. 

For $\omega=0.7 J_0$ and  edge excitation, we observe a localized mode at the array boundary in real space that is absent in the corresponding intensity distribution for the bulk excitation (compare Figs.~\ref{Fig4label} (e) and ~\ref{Fig4label} (g)). In Fourier space lowering the driving frequency leads to shrinkage of the Floquet Brillouin zone, which together with the aforementioned additional broadening hampers clear resolution of the band gap. In order to identify the edge mode we, therefore, concentrate on the brightest part of the spectrum in the $1^\mathrm{st}$ BZ. In case of the bulk excitation in Fig.~\ref{Fig4label}~(f) one sees the reduced intensity close to $k=0$ exactly at the position where the numerical calculations shown in Fig.~\ref{fig:theory2}~(f) predicted the occurrence of the band gap (indicated by the green arrow). Similar to the high frequency regime, the bands with a positive slope are predominantly populated. In contrast, in Fig.~\ref{Fig4label}~(h) the brightest line of the spectrum has no slope indicating the spatial localization of the corresponding mode.
Notably, the band gap at $k=0$ is no longer visible (see the magenta arrow), which suggests that this line should lie inside the gap. Comparing to numerical calculations in Fig.~\ref{fig:theory2}~(h),  we conclude that these spectral features can only correspond to the sought anomalous Floquet topological $\pi$-mode.

The anomalous Floquet topological $\pi$-mode can be also seen for $\omega=0.5 J_0$ and edge-excitation (real-space: Fig.~\ref{Fig4label} (k); Fourier-space: Fig.~\ref{Fig4label} (l)).
In the case of bulk-excitation, we also observe in real space a strongly suppressed spreading of the wavepacket (see Fig.~\ref{Fig4label} (i)) that corresponds in Fourier-space to nearly flat bands (see Fig.~\ref{Fig4label} (j)). 
Based on the discussion of the numerical calculations, we attribute this phenomenon to dynamic localization.

In conclusion, we presented a joint experimental and theoretical study of the periodically driven Su-Schrieffer-Heeger model implemented by evanescently coupled plasmonic waveguide arrays with periodically varying coupling constants. Calculations of the spectral density allow to predict different driving frequency regimes with distinct localization behavior. 
Our main finding is 
the spatially and spectrally resolved observation of the anomalous Floquet topological $\pi$-mode by leakage radiation microscopy  in the optical range.
Additionally, we demonstrated dynamic wave-packet localization in the bulk for suitable driving conditions such that the bulk bandwidth collapses.
The experimental findings agree well with the numerical calculations based on the Floquet theory.

We acknowledge financial support by the Deutsche Forschungsgemeinschaft through 
CRC/TR 185 (277625399) OSCAR and through the Cluster of Excellence ML4Q (90534769).

\bibliography{references}

\end{document}